\documentclass[]{article}

\def\ben{\begin{equation}}
\def\een{\end{equation}}
\def\bea{\begin{eqnarray}}
\def\eea{\end{eqnarray}}
\input amssym.def
\input amssym.tex
\begin{document}

\hfuzz=100pt
\title{Cosmological Evolution of the Rolling  Tachyon}
\author{G W Gibbons
\\
D.A.M.T.P.,
\\ C.M.S., \\ Cambridge University,
\\ Wilberforce Road,
\\ Cambridge CB3 0WA,
 \\ U.K.}
\maketitle

\begin{abstract}
The cosmological effects of the tachyon rolling down to its ground
state are discussed by coupling a  simple effective field theory
for the tachyon field to Einstein gravity. As the tachyon 
rolls down to the minimum of its potential the universe expands.
Depending upon initial conditions, the scale factor  may or may
not start off accelerating, but ultimately it ceases to do so and
the final flat spacetime is either static in the rest frame of the
tachyon  (if $k=0$) or ( if $k=-1$ ) given by the Milne model.

\end{abstract}

\vfill \eject

\section{Introduction} The reconciliation of fundamental theories such as M/String theory
with the basic facts of cosmology continues to present great
challenges. The most straightforward approach is to pass to an
effective field theory in which gravity is coupled to matter, for
instance in a supergravity theory. Because the theories are
formulated in higher dimensions, one must either  construct a
spontaneous compactification scenario or imagine a scheme in which
the universe is sort of 3-brane \cite{Gibbons-Wiltshire}. It seems
to be very difficult to construct models with $N=1$ supersymmetry
on the 3-brane \cite{Gibbons-Lambert}.  The trouble with
compactifications is that they come with associated massless
scalars and degenerate vacua: one must address the question of how
these evolve with time \cite{Gibbons-Townsend}. One must also
ensure that the resultant  time-variability of coupling constants
is compatible with observations.  One possibility is to give the
potentials a mass "by hand" such a way that Minkowski-space times
a Calabi-Yau is an attractor at late times,
\cite{Gibbons-Chapline} but this is {\it ad hoc} and ugly.

There is fairly good evidence from the BOOMERANG observations of
the  Cosmic Microwave Background that the scale factor of the
universe underwent a period of acceleration (so-called Primordial
Inflation)  at early times and from Type Ia super-novae that it
may also have been accelerating very recently if not today. (For a
recent review see \cite{Straumann}). It is quite difficult to get
accelerating universe out of pure supergravity theories
\cite{Gibbons, Maldacena-Nunez, Hull, Gibbons-Hull, Townsend,
Gates} although with super-matter, providing one gauges a suitable
axial current this is possible \cite{Freedman}. The problem is
that the axial gauging gives rise to anomalies \cite{Freedman2}.
It is possible that these  anomalies can be cancelled in staring
models with D-branes \cite{Kallosh}.

In recent years there has been great progress, particularly due to
Sen,  in our understanding of the role of the tachyon in String
Theory (see \cite{Sen1} for a recent account with references to
earlier work) . The basic idea is that the usual open string
vacuum is unstable but there exist a stable vacuum with zero
energy density which is stable. There is evidence that this state
is associated with the condensation of electric flux tubes of
closed strings (see \cite{Sen1, Sen}). These flux tubes described
successfully using an effective Born-Infeld action (see
\cite{Gibbons-Hori-Yi, Sen1, Sen3}  and references therein). This
success of effective action methods, together with the
difficulties of other approaches described the encourages one to
pursue this  further and to attempt a description of the cosmology
of tachyon rolling. Moreover not to take into account the effects
gravity during the process is  inconsistent, since it involves a
spatially uniform distribution of energy. It is the purpose of
this note to rectify this omission and initiate a study of tachyon
cosmology .

\section{The Rolling Tachyon} The tachyon of string theory may
described by effective field theory describing some sort of
tachyon condensate which in flat space has a Lagrangian density
\ben {\cal L} = -V(T) \sqrt{ 1+ \eta ^{\mu \nu}
\partial_\mu  T
\partial _\nu T}, \een where $T$ is the tachyon field, $V(T)$ is
the tachyon potential and $\eta_{\mu \nu} = {\rm diag}(-,+,+,\dots
)$ is the metric of Minkowski spacetime ( see \cite{Sen} for a
discussion with references to earlier work)  . The tachyon
potential $V(T)$ has a positive maximum at the origin and has a
minimum at $T=T_0$ where the potential vanishes. In \cite{Sen},
$T_0$ is taken to lie at infinity.  In Minkowski spacetime the
rolling down of the towards its minimum value is described by a
spatially homogeneous but time-dependent solution obtained  from
the Lagrangian density \ben {\cal L} = -V \sqrt{ 1-\dot T ^2 }.
\label{Lagrangian} \een During rolling the Hamiltonian density
\ben {\cal H} = {V(T) \over \sqrt {1-\dot T ^2 }}
\label{Hamiltonian}  \een has a constant value $E$. Thus \ben \dot
T = \sqrt { 1- {V^2(T)  \over E^2 }}\label{evolution}.\een As $T$
increases $V(T)$ decreases and $\dot T$ increases to attain its
maximum value of $1$ in infinite time as $T$ tends to infinity.
Note that as explained in \cite{Sen} the tachyon field behaves
like a fluid of positive energy density \ben \rho = {V(T) \over
\sqrt{1- \dot T ^2 }} \label{density} \een and negative pressure
\ben P= -V(T) \sqrt {1- \dot T ^2 }. \label{pressure} \een Thus
\ben P \rho = - V^2 (T) \een and \ben { P \over \rho} = w= -( 1-
\dot T^2 ), \een and therefore,  $-1 \le w \le 0$. Note that both
the Weak Energy Condition, $\rho >0$ and Dominant Energy
Condition, $\rho \ge |P|$ hold. However because \ben \rho + 3P =
-{2 V(T) \over \sqrt{1-\dot T^2 }} ( 1- { 3 \over 2} \dot T^2 )
\een the Strong Energy Condition fails to hold for small $|\dot
T|$ but does hold for large $|\dot T|$.

The discussion above has neglected the gravitational  field
generated by the tachyon condensate. To take it into account we
use the Lagrangian density \ben \sqrt{-g} \Bigl ( { R \over 16 \pi
G} - V(T) \sqrt {1 - g^{\mu \nu } \partial _\mu T \partial _\nu T
} \Bigr ), \een where $g_{\mu \nu}$  is the metric and $R$ its
scalar curvature. We shall work in $3+1$ spacetime dimensions and
assume that the metric has Friedman-Lemaitre-Robertson-Walker form
 \ben
ds^2 = -dt ^2 + a^2(t) d \Omega ^2 _k, \een where $a(t)$ is the
scale factor and $d \Omega^2 _k  $ is, locally at least,  the
metric on $S^3$ , ${\Bbb E } ^3$ or $ H^3$ according as $k=1,0,-1$
respectively. Note that in this model we have assumed that the
cosmological constant $\Lambda$ vanishes in the tachyon ground
state. The expressions (\ref{density} ) and ( \ref{pressure}) for
the density and pressure remain valid and thus the Friedman and
Raychaudhuri equations governing the evolution of the scale factor
are \ben { \dot a^ 2 \over a^2 } + { k \over a^2 } = { 8\pi G
\over 3}  { V(T) \over \sqrt { 1- \dot T^2 }}, \label{Friedman}
\een and

\ben { \ddot a \over a} = { 8\pi G \over 3} { V(T) \over \sqrt
{1-\dot T^2 }} \bigl ( 1- { 3 \over 2} \dot T ^2 \bigr ) .\label{Raychaudhuri}  \een

The Hamiltonian density of the tachyon field is no longer constant
because the tachyon Lagrangian density  is now explicitly time
dependent. Equations (\ref{Lagrangian})  and (\ref{Hamiltonian})
must  be replaced by  \ben {\cal L} = -a^3 V \sqrt{ 1-\dot T ^2 }.
\label{Lagrangian2} \een and  \ben {\cal H} = a^3 {V(T) \over
\sqrt {1-\dot T ^2 }} \label{Hamiltonian2} .\een The equation
\ben { d {\cal H} \over dt } = - { \partial  {\cal L} \over
\partial t} \label{timedependent} \een is formally equivalent to the
conservation of entropy of the fluid which reads \ben \dot \rho =
- { \dot a \over a}  (\rho + P).  \een Because  \ben \rho + P ={
\dot T^2 \over \sqrt{ 1- \dot T^2 } }, \een we have \ben { d \over
dt } \Bigl ( { V \over \sqrt {1- \dot T^2 }} \Bigr ) = - { 3 \dot
a \over a}  \Bigl ( { V \dot T^2 \over \sqrt{ 1- \dot T^2 }} \Bigr
). \een Thus the evolution equation (\ref{evolution}) remains
valid but the quantity $E$ is no longer constant but rather
decreases in time.

The course of cosmic evolution is now rather clear from these
equations. The tachyon field rolls down hill with an accelerated
motion  and the universe expands. It still follows from
(\ref{evolution} ) that as $T$ increases $V(T)$ increases but
since $E$ decreases, in principal $\dot T$ could decrease
but in any case $\dot T$ remains positive and so  $T$ increases
monotonically to attain its maximum value of 1.  From the Friedman
equation (\ref{Friedman}), it follows that if $k\le 0$, then $\dot
a $ will always be positive. This is because the Weak Energy
Condition holds, $\rho
>0$. From the Raychaudhuri equation (\ref{Raychaudhuri} ) one deduces that initially if
$ |\dot T | < { 2 \over 3} $  the scale factor initially
accelerates, $\ddot a >0 $ but eventually, once $\dot T$ exceeds
$\sqrt { 2\over 3}$  the acceleration will cease and deceleration will
set in. If the universe is flat, i.e. if $k=0$, then  ultimately
the scale factor will halt $a(t) \rightarrow {\rm constant}$. If
the universe has hyperbolic sections,  that is if $k=-1$, then
ultimately the scale factor increases linearly with time, $a
\rightarrow t$. In both cases the final state of the universe is
flat, the case $k=-1$ being the Milne model. In the case of
spherical sections, $k=+1$ re-collapse will take place. The
possibility of cosmic acceleration arises from the positive
potential $V(T)$ and should be contrasted with the situation in
pure supergravity theories for which the Strong Energy Condition
holds and cosmic acceleration is not possible \cite{Gibbons,
Maldacena-Nunez},  unless the internal space is non-compact
\cite{Hull, Gibbons-Hull}. The inclusion of supermatter  matter
may allow acceleration \cite{Townsend}.

Despite the violation of the Strong Energy condition one sees from
the Friedman equation (\ref{Friedman}) that in the cases $k\le 0$
the positivity of energy precludes the avoidance of singularities
in the past. If however $k=1$,  it is conceivable that, for
special initial conditions,  the scale factor might  pass through
a finite sequence of minima and maxima or even that periodic or
quasi-periodic solutions exist with an infinite sequence of maxima
and minima.

\section{Acknowlegement}

After submitting this paper to the archive I was told of
an earlier paper of Stephon Alexander \cite{SA} of which I was unaware
and which anticipated some
of the ideas discussed here, in terms of $D$-$\bar D$ annihilation.
The action adopted for the tachyon field is different from
that used in this paper. Another relevant pre-cursor
brought to my attention by Anupam Mazumdar and to which the same
remarks apply is  \cite{MPP}. I would like to thank them and
also Andrew Chamblin, Neil Lambert, Mohammad Garousi, Thanu
Padmanabhan, Nakoi Saskura and
Arkady  Tseytlin for helpful comments and pointng out typos
and small inaccuracies in the wording. 

\end{document}